\documentclass[a4paper]{jpconf}
\usepackage[table]{xcolor}
\usepackage{array,multirow,graphicx,amsmath,mathtools, %
  nicefrac,booktabs,tikz,adjustbox,hyperref,cite,wrapfig,etoolbox}
\hypersetup{colorlinks=true,linkcolor=blue,citecolor=cyan}
\newtoggle{preprint}

\toggletrue{preprint}

\graphicspath{{.}{./figures/}}

\iftoggle{preprint}{
\newcommand{\astronum}{{\sc Astronum-2017}}
\usepackage{tikzpagenodes,fancyhdr}
\pagestyle{fancy}\fancyhf{}
\fancyhead[L]{\astronum: Towards an exascale code for GRMHD on dynamical spacetimes}
\fancyhead[R]{\thepage}

}{} 

\makeatletter
\newcommand{\embedEQ}[1]{ {\CT@everycr{\the\everycr} #1 }}
\makeatother

\newcommand{\longNCP}[1]{ \parbox{4.0cm}{\embedEQ{#1}} }
\newcommand{\longSource}[1]{ \parbox{4.0cm}{\embedEQ{#1}} }

\newcommand{\param}[1]{\textcolor{red}{#1}}
\newcommand{\matter}[1]{\textcolor{blue}{#1}}

\definecolor{alpha}{HTML}{D1F2A5}
\definecolor{beta}{HTML}{FFE3EB}
\definecolor{gamma}{HTML}{FF9F80}
\definecolor{phi}{HTML}{FFDFAB}

\definecolor{Atilde}{HTML}{93DFB8} 
\definecolor{K}{HTML}{FFC8BA} 
\definecolor{Theta}{HTML}{E3AAD6} 
\definecolor{Gamma}{HTML}{B5D8EB} 
\definecolor{b}{HTML}{FFBDD8} 

\colorlet{A}{alpha}
\colorlet{B}{beta}
\colorlet{D}{gamma}
\colorlet{P}{phi}

\definecolor{varPP}{HTML}{DCEBFA}
\definecolor{varK0}{HTML}{FCF3CD}

\newcommand{\colorSymbol}[1]{ \cellcolor{#1} }
\newcommand{\colorNCP}[1]{ \cellcolor{#1!50!white} }
\newcommand{\colorSource}[1]{ \cellcolor{#1!60!white} }

\newcommand{\verticalrow}[2]{\parbox[t]{2mm}{\multirow{#1}{*}{\rotatebox[origin=c]{90}{#2}}}}

\newcommand{\Int}[2]{\int\limits_{\mathrlap{#1}}^{\mathrlap{#2}}}

\begin{document}
\title{Towards an exascale code for GRMHD on dynamical spacetimes}

\author{Sven K\"oppel}

\address{Frankfurt Institute for Advanced Studies (FIAS), Ruth-Moufang-Strasse 1, 60438 Frankfurt am Main, Germany; Institute for theoretical Physics (ITP), Goethe University Frankfurt, Max-von-Laue-Strasse 1, 60438 Frankfurt am Main, Germany}

\ead{koeppel@fias.uni-frankfurt.de}

\begin{abstract}
We present the first results on the construction of an exascale hyperbolic PDE engine (ExaHyPE), a code for the next generation of supercomputers with the objective to evolve dynamical spacetimes of black holes, neutron stars and binaries. We solve a novel first order formulation of Einstein field equations in the conformal and constraint damping Z4 formulation (CCZ4) coupled to ideal general relativistic magnetohydrodynamics (GRMHD), using divergence-cleaning. We adopt a novel communication-avoiding one-step ADER-DG scheme with an a-posteriori subcell finite volume limiter on adaptive spacetrees. Despite being only at its first stages, the code passes a number of tests in special and general relativity. 
\end{abstract}
%
%
\iftoggle{preprint}{
  \begin{tikzpicture}[remember picture,overlay]
    \node[align=center,text=black!50!white] at ([yshift=1em]current page text area.north)
     {\textsf{ Journal of Physics: Conference series, proceeding paper preprint:} \\ \textsf{\astronum{} --- the 12th International Conference on Numerical Modeling of Space Plasma Flows} };
  \end{tikzpicture}%
}{}
\section{Introduction}
Gravitational waves ejected by the merger of compact objects have been recently detected for both black hole-black hole mergers \cite{GW150914} as well as neutron star-neutron star mergers \cite{GW170817}. Both phenomena take place in the strong field regime of general relativity, described by Einstein field equations  and so far only accessible by numerical relativity. The general relativistic fluid description of matter allows to predict precisely the dynamics of single and binary neutron stars. Furthermode, electromagnetism plays an important role for the realistic description of the fate of the binary system. General relativistic (magneto)hydrodynamics---in short GR(M)HD---is a successful theory to describe these pheomena, also underlying the gravitational wave banks used for the recently detected NS-NS merger.

The coupled numerical evolution of Einsteins and the GRMHD equations is a comparatively young field of research, as the first successful evolution of a BH-BH system dates back only one decade \cite{Pretorius:2005gq}. Since then, many GRHD and GRMHD codes for solving coalescing neutron stars have been developed (for instance \cite{Baiotti04,Duez05MHD0,Anninos05c,Anton06,Giacomazzo:2007ti,Radice2012a,Radice2013b,Zanotti2015}, see also \cite{Font08,Marti2015} for a review).

In the time of gravitational wave multi-messenger astronomy, the quantiative and qualitative need for accurate numerical simulations is essential. In the same time, computational science is entering the era of \emph{exascale} computing, i.e. the first supercomputer to archive $10^{18}$ arithmetic operations per second is estimated to be built in the next three years. It is excepted that in order to reasonably exploit these resources, the next generation of numerical codes must satisfy tight constraints in locality, communication, storage, cache and energy efficiency \cite{exa1}.

\newpage\noindent
In this work, we present ingredients for such a code which solves
generic nonlinear hyperbolic first order balance laws,
\begin{equation}\label{eq.pde.base}
\partial_t Q_k + \partial_i F^i_k(Q)
+ B^{ij}_k(Q) \partial_i Q_j = S_k(Q)\,,
\end{equation}
with a state vector $\vec Q$, conserved fluxes $\vec F^i$, a
nonconservative part described by the matrix $\vec B^{ij}$ and
algebraic source terms $\vec S$. We present particular PDEs, defined by $\{\vec Q, \vec F^i, \vec B^{ij}, \vec S \}$,
as well as first results with actual implementations. The paper is structured as follows: In Section \ref{sec.pde} we present the PDE system, in Section \ref{sec.aderdg} we present the numerical scheme to solve this system, in Section \ref{sec.peano} we present the grid representation and communication, in Section \ref{sec.benchmarks} we present first benchmarks and results while in Section \ref{sec.summary} we give a summary and outlook of the next steps.

We work in geometric units (speed of light $c=1$ and gravitational constant $G=1$). Greek indices run from 0 to 3, Latin indices run from 1 to 3 and Einstein sum convention is applied over repeated incides. The signature of the metric tensor is assumed to be $(-,+,+,+)$.

\section{The Einstein-Maxwell-Euler system}\label{sec.pde}
Three theories which describe a wide range of astrophysical phenomenae are
\begin{alignat}{3}
\label{eq.intro.efe}
&\text{Einstein equations}
\quad\quad
& R g_{\mu\nu} + R_{\mu\nu} &= 8\pi G ~ T_{\mu\nu}
\,,
\\
\label{eq.intro.maxwell}
&\text{Maxwell equations}
\quad\quad
&\nabla_\mu ^* F^{\mu\nu} &=0
~~\text{and}~~
\nabla^\mu F_{\mu\nu} = 0
\,,
\\
\label{eq.intro.hydro}
&\text{Euler equations}
\quad\quad
& \nabla^\mu T_{\mu\nu} &= 0
~~\text{and}~~
\nabla_\mu (\rho u^\mu) = 0
\,.
\end{alignat}
with Ricci scalar $R$, metric $g_{\mu\nu}$, energy momentum tensor $T_{\mu\nu}$, Faraday tensor $F^{\mu\nu}$, dual Faraday tensor $^* F_{\mu\nu}$, rest mass density $\rho$ and fluid velocity $u^\mu$. In the following, 3+1 initial value formulations in the language of \eqref{eq.pde.base} are presented.
\subsection {A first order formulation of Einstein equations: FO-CCZ4}
The formulation of Einstein equations used in this work builds upon successors of the original ADM Cauchy initial value formulation of Einsteins equations \cite{Arnowitt62unfindable} which introduces the 16~dynamical fields lapse $\alpha$, shift $\beta^i$, 3-metric $\gamma_{ij}$ and 3-extrinsic curvature $K_{ij}$. The ADM equations are not hyperbolic \cite{Baumgarte2010, Gourgoulhon2012, Rezzolla_book:2013}. Two successors of the ADM equations are the BSSNOK formulation \cite{Shibata95,Baumgarte99} and the class of Z4 formulations \cite{Bona:2003fj,Alic:2009}. Both approaches were unified in the conformal and constraint-damping CCZ4 formulation \cite{Alic:2011a,Alic2013} which was recently rewritten in a first order formulation in space and time \cite{ADERCCZ4} which we subsequently refer to as FO-CCZ4. The full derivation and discussion of these equations including an eigenstructure analysis can be found in \cite{ADERCCZ4} where also strong hyperboliciy for the system in certain gauges is proven.

In order to briefly outline the equations, we introduce the conformal factor $\phi=(\det \gamma_{ij})^{-1/6}$ and subsequently $\tilde \gamma_{ij}=\phi^2 \gamma_{ij}$. We also define a conformal and trace-free extrinsic curvature tensor $\tilde A_{ij}=\phi^2 (K_{ij} - \nicefrac 13~ K \gamma_{ij})$ where it's trace $K=K_{ij} \gamma^{ij}$ has been seperated. Furthermore, we define the Christoffel variable contraction $\tilde \Gamma^i = \tilde \gamma^{ij} \tilde \Gamma^i_{jk}$ as well as $\hat \Gamma^i=\tilde \Gamma^i + 2 \tilde \gamma^{ij} Z_j$. Recall that $Z_\mu=(\Theta,Z_i)$ is the vector field which measures the distance to an analytical solution of Einsteins equations with constraint $Z_\mu=0$. For the first-order formulation, we introduce the auxilliary vectors $A_i = \partial_i \ln \alpha$ and $P_i=\partial_i \ln \phi$ as well as tensors $B^i_k=\partial_k \beta^i$ and $D_{kij}=\nicefrac 12 \partial_k \tilde \gamma_{kj}$. Based on these quantities, the FO-CCZ4 system is given in Table \ref{table.pde.foccz4} where system parameters $(\tilde \tau, s,f,e,\kappa_{1,2,3},c,\eta,\mu)$ and slicing condition $g(\alpha)$ are highlighted in red whereas matter contributions (cf. Section \ref{sec.grmhd}) are highlighted in blue. The state vector $Q_i$ collects 59 evolved variables. For details on the imposed constraint equations, applied Bona-Mass\'o gauge conditions and meaning of the parameters we refer again to \cite{ADERCCZ4}.

\begin{table}
\caption{FOCCZ4 system $\partial_t Q + B(Q)\nabla Q = S(Q)$}\label{table.pde.foccz4}
\centering
\begin{adjustbox}{center,scale=0.8}
\begingroup 
\setlength{\abovedisplayskip}{0pt}
\setlength{\belowdisplayskip}{0pt}
\setlength{\abovedisplayshortskip}{0pt}
\setlength{\belowdisplayshortskip}{0pt}
\let\OLDdisplaystyle\displaystyle
\let\displaystyle\textstyle

\begin{tabular}{llll}
\br
& $Q_i$ & NCP$_{Q_i}$: Nonconservative product $B \nabla Q$ & $S_{Q_i}$: Algebraic source $S$ \\
\mr
\verticalrow{4}{ODE-ADM}
& \colorSymbol{alpha} $\ln \alpha$ 
& \colorNCP{alpha} $~0$
& \colorSource{alpha} ${ \beta^k A_k } - \alpha \param{g}(\alpha) ( K - K_0 - 2\Theta {c} ) $ \\
& \colorSymbol{beta} $\beta^i$
& \colorNCP{beta} $~0$ 
& \colorSource{beta} $\param s \beta^k B_k^i + \param s \, \param f \, b^i$  \\
& \colorSymbol{gamma} $\tilde \gamma_{ij}$
& \colorNCP{gamma} $~0$
& \colorSource{gamma} $
{\beta^k 2 D_{kij} + \tilde\gamma_{ki} B_{j}^k  + \tilde\gamma_{kj} B_{i}^k - \nicefrac{2}{3}\tilde\gamma_{ij} B_k^k }
	- 2\alpha \big( \tilde A_{ij} - {\nicefrac{1}{3}~ \tilde \gamma_{ij} \textnormal{tr}{\tilde A} } \big)
  - { \nicefrac{1}{\param{\tilde\tau}} ( \tilde{\gamma} -1 ) \, \tilde{\gamma}_{ij}}
$
\\
& \colorSymbol{phi} $\ln \phi$
& \colorNCP{phi} $~0$
& \colorSource{phi} ${ \beta^k P_k } + \nicefrac{1}{3} \left( \param \alpha K - {B_k^k} \right)$  \\
\midrule
\verticalrow{5}{SO-CCZ4}
& \colorSymbol{Atilde} $\tilde A_{ij}$
& \colorNCP{Atilde} \longNCP{\begin{align*}
&-\beta^k \partial_k\tilde A_{ij}
+ \nicefrac{1}{3}~ \tilde\gamma_{ij}  \left( -\nabla^k \nabla_k  \alpha  +  \alpha R +  2  \alpha \nabla_k Z^k  \right)
\\&
-  \phi^2 \left(  -\nabla_i\nabla_j \alpha +  \alpha R_{ij} + \alpha \nabla_i Z_j  +  \alpha \nabla_j Z_i \right)
\end{align*}}
& \colorSource{Atilde} \longSource{\begin{align*}
&{ \tilde A_{ki} B_j^k + \tilde A_{kj} B_i^k - \nicefrac{2}{3}\,\tilde A_{ij} B_k^k }~
\matter{
-\phi^4 8 \pi \left( {S_{ij}} - \nicefrac{1}{3}\, \tau \tilde g_{ij} \right) }
\\&
+ \alpha \tilde A_{ij}(K - 2 \Theta {c} )  - 2 \alpha\tilde A_{il} \tilde\gamma^{lm} \tilde A_{mj}  - \nicefrac 1{\param{\tilde\tau}} \, \tilde{\gamma}_{ij} \, \textnormal{tr}{\tilde A}    
\end{align*}} \\
& \colorSymbol{K} $K$
& \colorNCP{K} $~ - \beta^k \partial_k K  + \nabla^i \nabla_i \alpha - \alpha( R + 2 \nabla_i Z^i)$
& \colorSource{K} $\alpha K (K - 2 \, \Theta \,\param{c} ) - 3\alpha\param{\kappa_1}(1+\param{\kappa_2})\Theta + \matter{4\pi {(S-3\tau)}} $ \\
& \colorSymbol{Theta} $\Theta$
& \colorNCP{Theta} $~ - \beta^k\partial_k\Theta  -  \nicefrac{1}{2}~\alpha {\param e^2} ( R + 2 \nabla_i Z^i)$
& \colorSource{Theta} $ \nicefrac{1}{2}~\alpha {\param e^2} ( \nicefrac{2}{3} K^2 - \tilde{A}_{ij} \tilde{A}^{ij} ) - \alpha \Theta K \param{c} - {Z^i \alpha A_i} - \alpha\param{\kappa_1}(2+ \param{\kappa_2})\Theta~ \matter{- 8\pi \alpha {\tau}}$  \\
& \colorSymbol{Gamma} $\hat \Gamma^i$
& \colorNCP{Gamma} \longNCP{\begin{align*}
&- \beta^k \partial_k \hat \Gamma^i + \nicefrac{4}{3}~ \alpha \tilde{\gamma}^{ij} \partial_j K  - 2 \alpha \tilde{\gamma}^{ki} \partial_k \Theta \\
&- \param s\tilde{\gamma}^{kl} \partial_{(k} B_{l)}^i
- \nicefrac{\param s}{3}~ \tilde{\gamma}^{ik}  \partial_{(k} B_{l)}^l - { \param s 2 \alpha \tilde{\gamma}^{ik}  \tilde{\gamma}^{nm} \partial_k \tilde{A}_{nm}   }
\end{align*}}
& \colorSource{Gamma} \longSource{\begin{align*}
&{ \nicefrac{2}{3} \tilde{\Gamma}^i B_k^k - \tilde{\Gamma}^k B_k^i  } +
       2 \alpha ( \tilde{\Gamma}^i_{jk} \tilde{A}^{jk} - 3 \tilde{A}^{ij} P_j ) - 
       2 \alpha \tilde{\gamma}^{ki} \left( \Theta A_k + \nicefrac{2}{3} K Z_k \right)
       \matter{ - 16\pi \alpha \tilde{\gamma}^{ij} {S_j} }
\\&	-	 2 \alpha \tilde{A}^{ij} A_j 
	   - 4\param s \alpha \tilde{\gamma}^{ik} D_k^{nm} \tilde{A}_{nm}
 + 2\param{\kappa_3} \left( \nicefrac{2}{3}~ \tilde{\gamma}^{ij} Z_j B_k^k - \tilde{\gamma}^{jk} Z_j B_k^i \right)
- 2 \alpha \param{\kappa_1} \tilde{\gamma}^{ij} Z_j 
\end{align*}} \\
& \colorSymbol{b} $b^i$
& \colorNCP{b} $~ - \param s \beta^k \partial_k b^i $
& \colorSource{b} $\param s (  \partial_t \hat\Gamma^i - \beta^k \partial_k \hat \Gamma^i - \param \eta b^i )$
\\
\midrule
\verticalrow{5}{FO-CCZ4}
& \colorSymbol{A} $A_k$
& \colorNCP{A} \longNCP{\begin{align*}
&- {\beta^l \partial_l A_k} + \alpha \param g(\alpha) \left( \partial_k K - \partial_k K_0 - 2 \param c \partial_k \Theta \right) \\
&+ {\param s \, \alpha \, \param g(\alpha) \tilde{\gamma}^{nm} \partial_k \tilde{A}_{nm} }
\end{align*}}
& \colorSource{A} \longSource{\begin{align*}
&- {\param s \, \alpha \, \param g(\alpha) \partial_k \tilde{\gamma}^{nm} \tilde{A}_{nm} } \\
&-\alpha A_k \left( K - K_0 - 2 \Theta \param c \right) \left( \param g(\alpha) + \alpha  \param g'(\alpha)  \right) + B_k^l ~A_{l}
\end{align*}} \\
& \colorSymbol{B} $B_k^i$
& \colorNCP{B} \longNCP{\begin{align*}
&- \param s\beta^l \partial_l B_k^i - \param s\big(  f \partial_k b^i - { \param \mu \, \tilde{\gamma}^{ij} \left( \partial_k P_j - \partial_j P_k \right) } \\
& + \param \mu \, \tilde{\gamma}^{ij} \tilde{\gamma}^{nl} \left( \partial_k D_{ljn} - \partial_l D_{kjn} \right)  \big)
\end{align*}}
& \colorSource{B} $ B^l_k~B^i_l $\\
& \colorSymbol{D} $D_{kij}$
& \colorNCP{D} \longNCP{\begin{align*}
& - {\beta^l \partial_l D_{kij}}  
         - \nicefrac{\param s}{2}~ \tilde{\gamma}_{mi} \partial_{(k} {B}_{j)}^m
         - \nicefrac{\param s}{2}~ \tilde{\gamma}_{mj} \partial_{(k} {B}_{i)}^m
\\&		 + \nicefrac{\param s}{3}~ \tilde{\gamma}_{ij} \partial_{(k} {B}_{m)}^m   		 +  \alpha \partial_k \tilde{A}_{ij}
		-  {\nicefrac{1}{3}~ \alpha \tilde{\gamma}_{ij} \tilde{\gamma}^{nm} \partial_k \tilde{A}_{nm} } 
\end{align*}} 
& \colorSource{D} \longSource{\begin{align*}
& B_k^l D_{lij} + B_j^l D_{kli} + B_i^l D_{klj} - \nicefrac{2}{3}~ B_l^l D_{kij} + { \nicefrac{1}{3}~ \alpha \tilde{\gamma}_{ij} \partial_k \tilde{\gamma}^{nm} \tilde{A}_{nm} } \\
& - \alpha A_k ( \tilde{A}_{ij} - \nicefrac{1}{3}~ \tilde{\gamma}_{ij} \textnormal{tr} \tilde{A} )
\end{align*}}  \\
& \colorSymbol{P} $P_k$
& \colorNCP{P} \longNCP{\begin{align*}
{\beta^l \partial_l P_{k} - \nicefrac{1}{3} ~ \alpha \partial_k K
+ \nicefrac{1}{3} ~ \partial_{(k} {B}_{i)}^i  } - \nicefrac{\param s}{3} ~ \alpha \tilde{\gamma}^{nm} \partial_k \tilde{A}_{nm}
\end{align*}}
& \colorSource{P}
$\nicefrac{1}{3} ~ \alpha A_k K + B_k^l P_l + \nicefrac{\param s}{3} ~ \alpha \partial_k \tilde{\gamma}^{nm} \tilde{A}_{nm}$
\\
\bottomrule
\end{tabular}
\endgroup 
\end{adjustbox}
\end{table} 

\begin{wraptable}{r}{85mm}
\caption{GRMHD system components}\label{table.pde.grmhd}
\begin{center}
\begin{adjustbox}{center,scale=0.8}
\begin{tabular}{llll}
\br
$Q_i$ & Conserved Flux $F^i$ & Nonconservative product $B \nabla Q$  \\
\mr
$D$ & $w^i D$ & $0$ \\
$S_j$ & $\alpha W^i_j - \beta^i S_j$ 
& $E \partial_j \alpha -\frac \alpha2 S^{lm} \partial_j \gamma_{lm} - S_k \partial_j \beta^k$ \\
$\tau$ & $\alpha (S^i - v^i D) - \beta^i \tau$
& $S^i \partial_i \alpha - \alpha S^{ij} K_{ij}$ \\
$B^j$ & $w^i B^j - v^j B^i - B^i \beta^j$ 
& $B^i \partial_i \beta^j + \alpha \gamma^{ij} \partial_i \phi $ \\
$\phi$ & $\phi \beta^i - \alpha B^i$
& $\phi \partial_i \beta^i + \frac 12 \phi \gamma^{ij}
\beta^k \partial_k \gamma_{ij} - B^i \partial_i \alpha$ \\
\br
\end{tabular}
\end{adjustbox}
\end{center}
\vspace*{-0.5cm} 
\end{wraptable}

\subsection{The GRMHD system}\label{sec.grmhd}

The GRMHD system describes general relativistic hydrodynamics coupled to the magnetohydrodynamic approximation of Maxwell
theory where magnetic field lines are advected with the fluid and the
electric field is given by $\vec E = - \vec v \times \vec B$.
Thus, the Maxwell equations reduce to an evolution equation for the
magnetic field as well as the constraint equation $\nabla \cdot \vec B=0$. Violations are treated with a cleaning approach, i.e. we evolve an additional scalar field $\phi$ to transport constraint violations (no damping is applied). The PDE system is given in Table~\ref{table.pde.grmhd}, where $D=W\rho$ is the conserved density which is related to the rest mass density $\rho$ by the Lorentz factor $W=(1-v_i v^i)^{-1/2}$ with fluid velocity $v^i$.  The conserved momentum is given by $S_j=\rho h W^2 v_j$. We also evolve the rescaled energy density $\tau=E-D=\rho h W^2 - p - \rho W$ where $E$ is the energy density measured by the Eulerian observer and $p$ is the fluid pressure. $B^i$ represents the magnetic field and $w^i=\alpha v^i - \beta^i$ the advection velocity relative to the coordinates, also refered to as transport velocity. The GRMHD 3-energy-momentum tensor is given by $S^{ij}=S^i v^j + (p-T^k_k/2) - T^{ij}$ with $T^{ij}=B^i B^j/W^2 + (B^i v^j)(B^k v_k)$ the Maxwell 3-energy momentum tensor. The primitive recovery follows \cite{NGMD2006,ZannaZanotti,BHAC2017}. Note that all fields are rescaled to a tensor density by $\sqrt g \equiv \sqrt{\det g_{ij}}$ and the actual PDE therefore reads
\begin{equation}
\partial_t \sqrt{g} Q_i + \nabla_j \sqrt{g} F^j(Q) + \sqrt{g} B \nabla Q = 0
\,.
\end{equation}
For a comprehensive discussion of this system in context of our ADER-DG approach (Section~\ref{sec.aderdg}), we refer to our publication \cite{fambri17}.
\clearpage
\section{Communication-avoiding ADER-DG scheme with a subcell limiter}\label{sec.aderdg}

We adopt a path-conservative arbitrary high order derivative discontinous Galerkin (ADER-DG) scheme
with a finite volume (FV) limiting approach,
adaptive mesh refinement (AMR) and local timestepping. This scheme
was already applied on several nonrelativistic problems \cite{Zanotti2015,Zanotti2015c,Zanotti2015d,ADERDGVisc} and recently formulated for GRMHD \cite{fambri17}.

The computational (spatial) domain $\Omega$ is covered by Cartesian nonoverlapping \emph{cells} $\Omega_i$ (also refered to as \emph{patches} or \emph{elements}). The analytic DG solution $u_h(x, t^n)=\sum_l \hat u_{il}^n \Phi_l(x)$ is written in an orthogonal basis $\Phi_l(x)$ in each
cell. We choose the Euler-Legendre polynomial basis at order $N$ and thus store $N+1$ degrees of freedom (DOF) on a non-uniform nodal \emph{subgrid}. The PDE system~\eqref{eq.pde.base} is multiplied by these spatial test functions and written in weak integral form as
\begin{equation}\label{eq.aderdg.start}
0 =
\Int{t^n}{t^{n+1}} \mathrm{d}t
\int\limits_{\Omega_i}  \mathrm{d}x
\left(
\frac{\partial Q}{\partial t}
+ \frac{\partial F^i(Q)}{\partial x^i}
+ B^{ij} \frac{\partial Q_j}{\partial x^i}
- S(Q)
\right) \Phi_k 
\,.
\end{equation}

\subsection{Space-time predictor}
In order to archieve element-locality, the scheme \eqref{eq.aderdg.start} is seperated in an element-local space-time \emph{predictor} and a \emph{corrector} step where boundary face values are communicated,
\begin{equation}
0=
\left( \hat u^{n+1}_{kl} - \hat u^{n}_{kl} \right) M_{kl}
+
\Int{t^n}{\mathrlap {t^{n+1}}}
\mathrm{d}t~
\Big\{
\Int{\partial\Omega_i}{}
\vec{n} \mathrm{d}S~
\Phi_k 
\mathcal{J}(q^-_h,q^+_h)
+
\Int{\Omega_i \setminus \partial \Omega_i}{}
\mathrm{d}x~\Phi_k
\left[
B(q_h) \nabla q_h - S
\right]
\Big\}
\,.
\end{equation}
Here, the $q_h$ is the element-local predictor solution and $M_{kl}=\int_{\Omega_i} \mathrm{d}x\, \Phi_k \Phi_l$ the diagonal element mass matrix. The surface integral collects the conservative and nonconservative jump terms $\mathcal{J}$ between cells and is subject of an approximate Riemann solver, we apply a path-conservative HLLEM method \cite{NCP_HLLEM} which takes the nonconservative terms into account.

The predictor solution $q_h(x,t)$ is approximated in time from a known solution without considering the interaction with the neighbouring cells. We thus solve a \emph{local} Cauchy problem inside the cell,

\begin{align}\label{eq.aderdg.stp}
&
\Int{t^n}{t^{n+1}} \mathrm{d}t
\Int{\Omega_i}{} \mathrm{d}x \,
\Theta_k(x,t)
\left( S(q_h) - B^{ij} \nabla q_h - \nabla F(q_h) \right)
=
\Int{t^n}{t^{n+1}} \mathrm{d}t
\Int{\Omega_i}{} \mathrm{d}x \,
\Theta_k(x,t) \frac{\partial q_h}{\partial t}
\\
&=
\Int{\Omega_i}{} \mathrm{d}x
\Theta_k(x,t^{n+1}) \left(
q_h(x,t^{n+1}) - u_h(x,t^n) \right)
- \Int{t^n}{t^{n+1}} \mathrm{d}t \Int{\Omega_i}{} \mathrm{d}x
\frac{\partial \Theta_k}{\partial t} 
q_h(x,t)
\,,
\end{align}
which was rewritten by partial integration in order to archieve a fixed-point problem which can easily be solved with fast converging algorithms \cite{Dumbser2008,ADERNC}.

\subsection{Subcell limiter}
In case of discontinuities in the solution, the ADER-DG scheme can produce spurious oscillations (Gibbs phenomena) or even unphysical solutions. We encounter these problems by switching to a robust finite volume scheme on a regular subgrid. This switch is either done \emph{a priori} on a purely geometrical criterion (e.g. static puncture position) or \emph{a posteriori}, \emph{after} computing a DG timestep. In the later case, several criteria are applied to ensure the correctness of the solution: Mathematical admissibility criteria (the occurance of floating point errors), physical admissibility criteria (e.g. $D>0$, $\vec v^{\,2}<1$, $\rho>0$ in the GRMHD equations or $\alpha<\alpha_0$ in CCZ4) as well as an heuristic discrete maximum principle (DMP) which checks whether all cell solutions are below the maxima and above the minima of the adjacent cells.

By choosing a subgrid size of $M=2N+1$ cells per axis, where $N$ is the order of the DG polynomial, the number of subcells is maximized without breaking the CFL condition which would make the limited cells evolve with smaller timesteps than untroubled DG cells. We can not only preserve all $N+1$ DOF per axis of a troubled DG cell, the limiter also acts as a natrual mesh refinement on its own.

In principle, any FV scheme can be used. We implemented an ADER-TVD  \cite{fambri17} and ADER-WENO scheme \cite{ADERCCZ4} but also, for communication avoidance, a 1st order Godunov scheme and a 2nd order MUSCL-Hancock scheme \cite{exahype-review}. The limiter is extensively described in the references \cite{ADERCCZ4,fambri17,exahype-review}. In short, it completely replaces the DG scheme in the troubled cell. In order to properly communicate the boundary values, a DOF prolongation of DG$\to$FV is also neccessary in the adjacent cells. After each timestep, the restriction FV$\to$DG takes place to restore the DG polynomial.

\section{Exascale architecture}\label{sec.peano}
It is assumed that many assumptions of contemporary high performance computing no longer hold at the exascale. For instance, due to the heterogenity and complex, error-prone and fault-tolerant computer architectures, equal work distribution (i.e. naive Cartesian slicing which minimizes the allocation surface) is no longer assumed to lead to balanced computing time. Similarly, massive parallel scalability will become even more important, and thus the communication overhead has to be reduced at any price \cite{exa1}.

\subsection{Seperation of physics and machinery: Loose of control}

Cactus and the Einstein Toolkit \cite{loeffler_2011_et,Goodale02a} are examples of mature frameworks for PDE solvers. While they focus on abstracting the distributed memory parallelization (MPI), the advent of manycore processors requires hybrid codes (shared memory parallelization, e.g. OpenMP or TBB) and diffused the clear seperation of physics and computational infrastructure \cite{loeffler2013,schnetter14}. In contrast, the novel ExaHyPE code forces the users to step back from implementing random access schemes as it prescribes both the data layout, programming workflow as well as the major algorithmic steps \cite{exahype-review,exahype-web,exahype-guidebook}. Primarily, the physicist as a user \emph{declares} the PDE system without describing its control flow. 

\subsection{Spacetime AMR by Peano}

We use the Peano framework \cite{Weinzierl15} which implements
tree-structured Cartesian meshes with adaptive mesh refinement
(AMR). The name indicates that the Peano space filling curve is
employed for storage locality \cite{bader-sfc}. The code also supports
Hybrid (MPI and TBB) parallelization with dynamic load balancing.

Peano implements the inversion of control paradigm where users give up control
of the program workflow in order to profit from high-level task restructuring.
Peano defines the program workflow in terms of two coupled finite state machines, one for the grid traversal and one for the major steps of the PDE solver. Peano is one of the building blocks of the exascale hyperbolic PDE engine ExaHyPE \cite{exahype-web}. In ExaHyPE, the concrete numerical scheme (e.g. the particular Riemann solver used) is exchangable and delegated to modular \emph{kernels} which can be tailored to a specific PDE.

\section{Benchmarks and Results}\label{sec.benchmarks}

This section collects seperated results from vacuum spacetime \cite{ADERCCZ4}, GRMHD in fixed background Cowling approximation \cite{fambri17}, and dedicated nonrelativistic tests with the limiter in the Peano framework \cite{exahype-review}. All figures are taken from the respective publications. The evolved fluids follow an ideal gas equation of state. 

\subsection{Nonrelativistic benchmarks}
\begin{figure}[h]
\begin{minipage}[t]{0.25\textwidth}
\includegraphics[width=\textwidth]{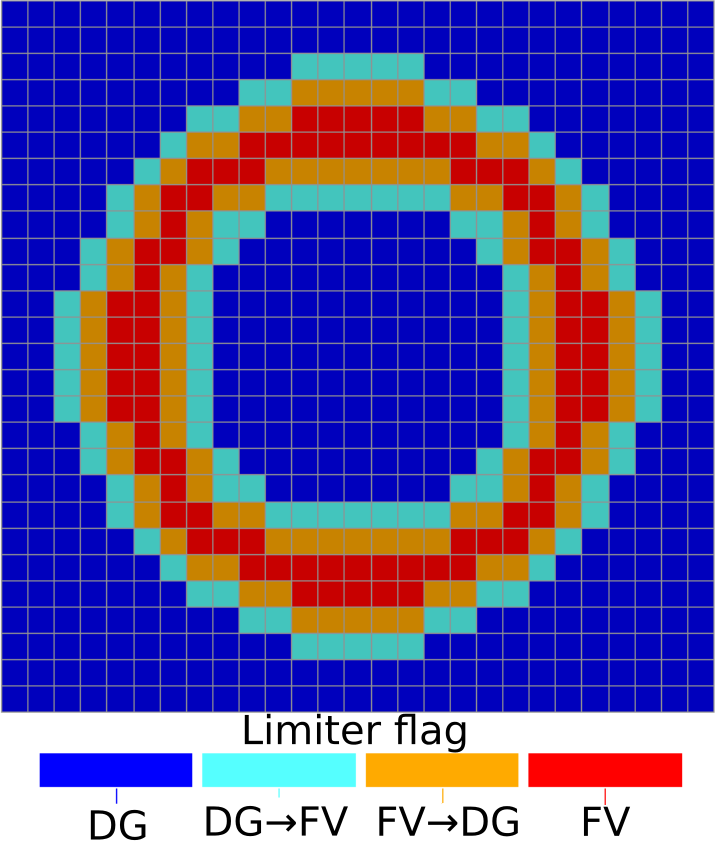}
\caption{\label{img.bench.limiter}Radial explosion problem with clearly visible cell status.}
\end{minipage}\hspace{2pc}%
\begin{minipage}[t]{0.40\textwidth}
\includegraphics[width=\textwidth]{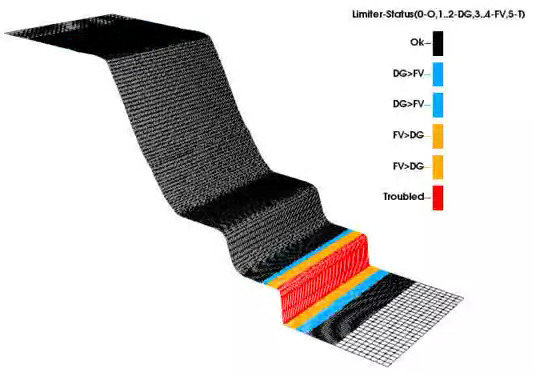}
\caption{\label{img.bench.shock}2D Sod shock tube. The rarefraction wave and shock front are computed with ADER-DG, only the contact discontinuity is limited.}
\end{minipage} \hspace{2pc}%
\begin{minipage}[t]{0.25\textwidth}
\includegraphics[width=\textwidth]{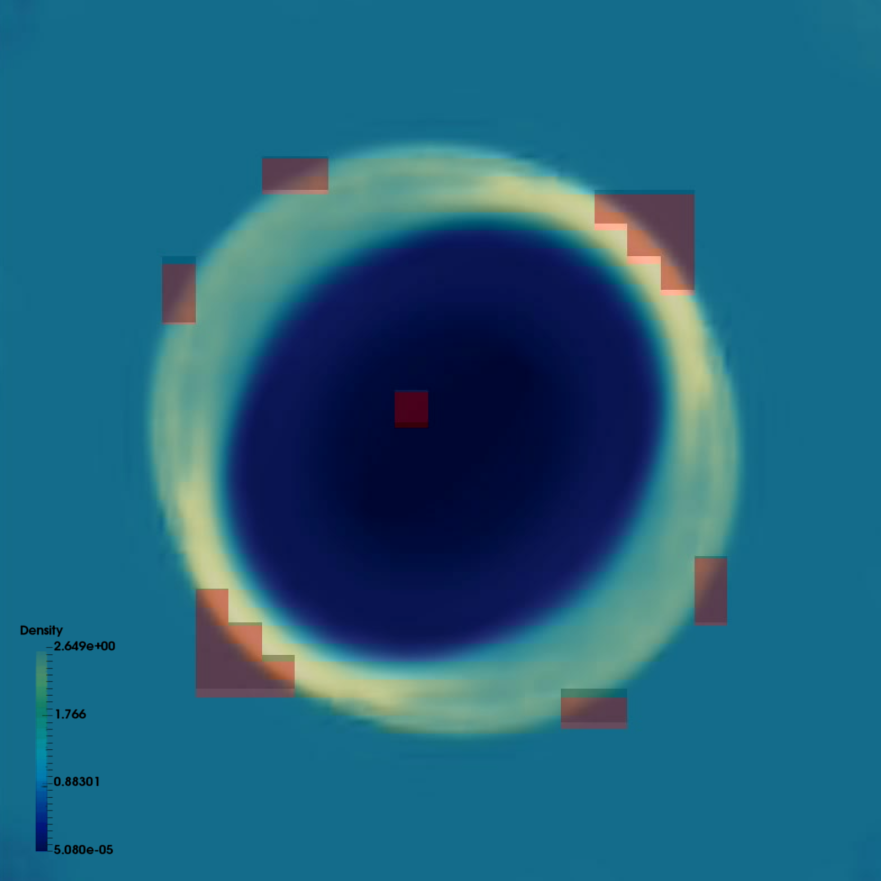}
\caption{\label{img.bench.rotor}MHD Rotor, poorly resolved: Coincidental limiting due to heuristics.}
\end{minipage} 
\end{figure}

Figure \ref{img.bench.limiter} visualizes the Limiter functionality on an Eulerian explosion problem (spherical Sod shock tube \cite{sod,toro-book}). Figure 
\ref{img.bench.shock} shows a similar but non-spherical Sod shock tube. Both figures highlight the transition steps between limited (troubled) cells and unlimited cells. Figure \ref{img.bench.rotor} shows the 2D SRMHD rotor problem \cite{Balsara1999b} in a poor resolution, cf. \cite[fig. 4]{Zanotti2015d} for a high resolution. Since the limiter acts only per-cell, a minimum resolution is neccessary to properly work.
\subsection{Cowling GRHD benchmarks}
\begin{figure}[h]
\begin{minipage}{0.35\textwidth}
\includegraphics[width=\textwidth]{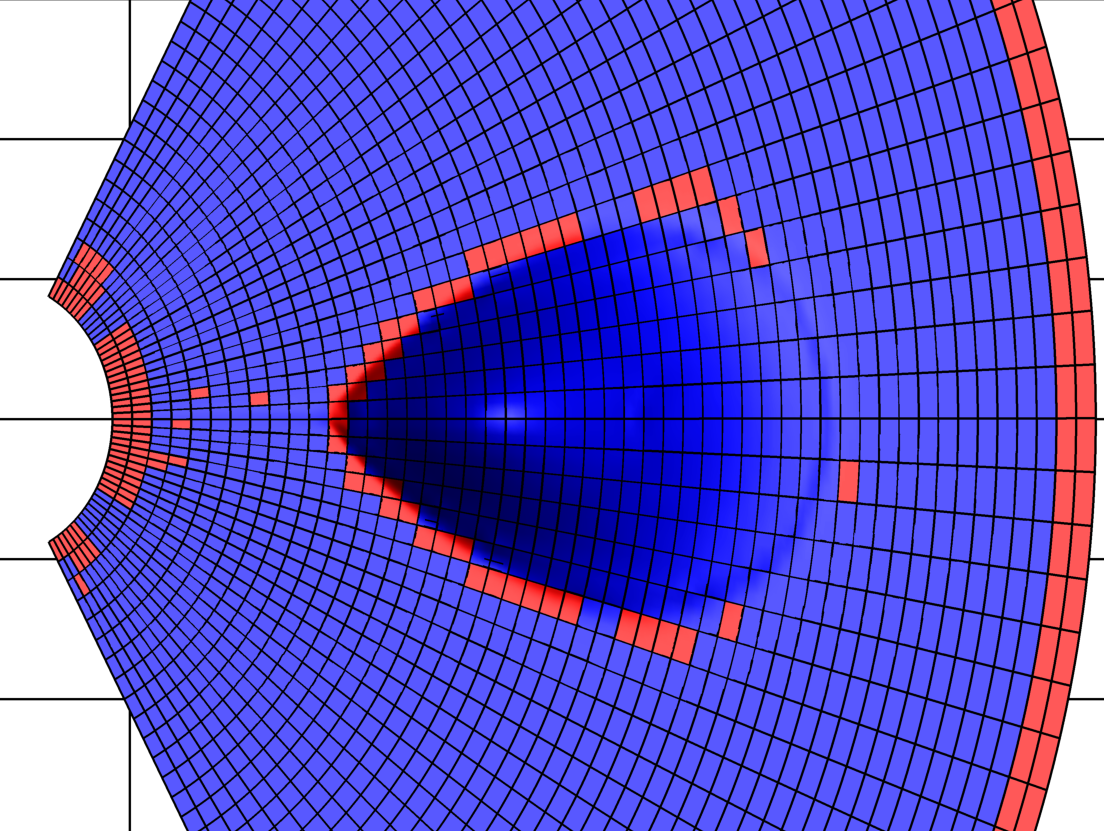}
\caption{\label{img.bench.torus}2D thick torus in polar coordinates. The red cells indicate limited cells.}
\end{minipage}\hspace{2pc}%
\begin{minipage}{0.65\textwidth}
\includegraphics[width=\textwidth]{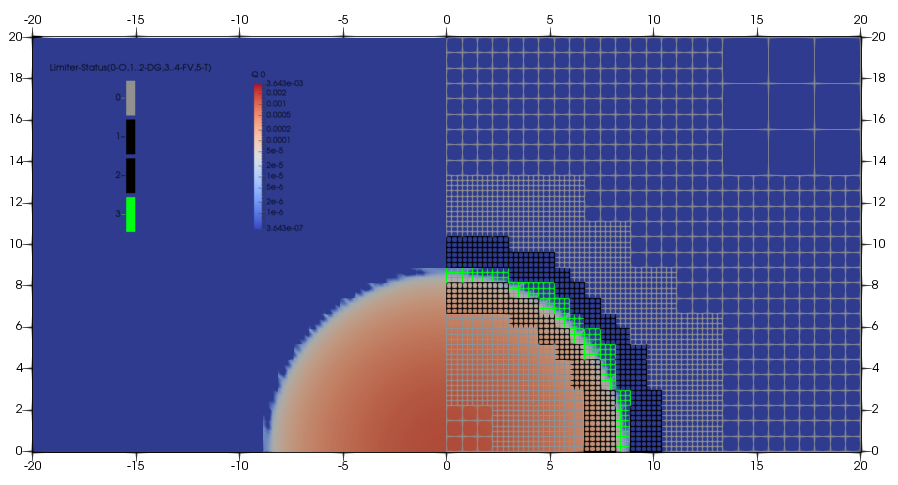}
\caption{\label{img.bench.ns}3D cut of a stable neutron star. Green cells indicate limited cells on a 3-level AMR grid.}
\end{minipage} 
\end{figure}

Figure \ref{img.bench.torus} shows the Limiter active in a 2D thick torus
\cite{Font02a} solution after $t=1000M$. Figure \ref{img.bench.ns} shows
the grid around a TOV \cite{Tolman39} neutron star surface. Both figures show the color encoded rest mass density profile. The limiter always activates at the sharp crossover from matter to vacuum (atmosphere). 

\subsection{Vacuum spacetime benchmarks}
\begin{figure}[h]
\begin{minipage}{0.55\textwidth}
\includegraphics[width=0.9\textwidth]{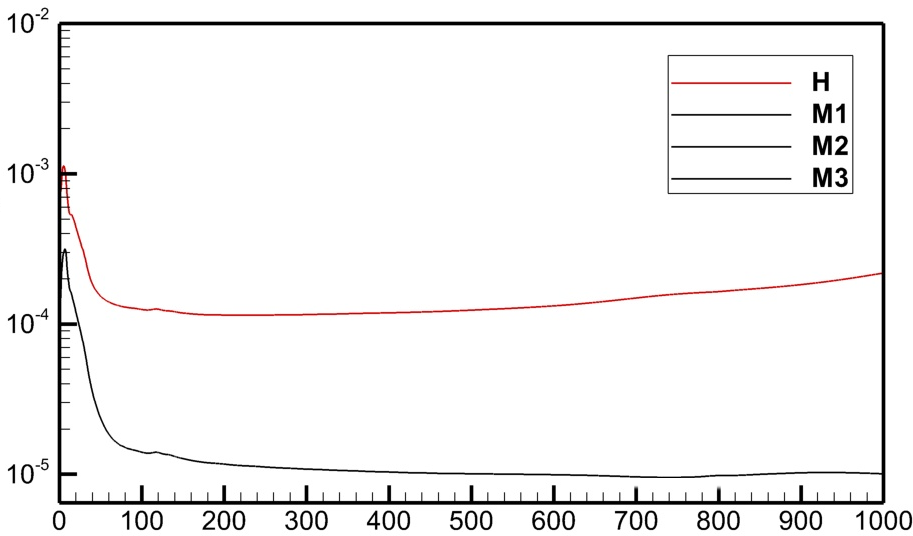}
\caption{\label{img.bench.bh}Errors in the black hole spacetime evolution up to $t=1000M$.}
\end{minipage}\hspace{2pc}%
\begin{minipage}{0.35\textwidth}
\includegraphics[width=\textwidth]{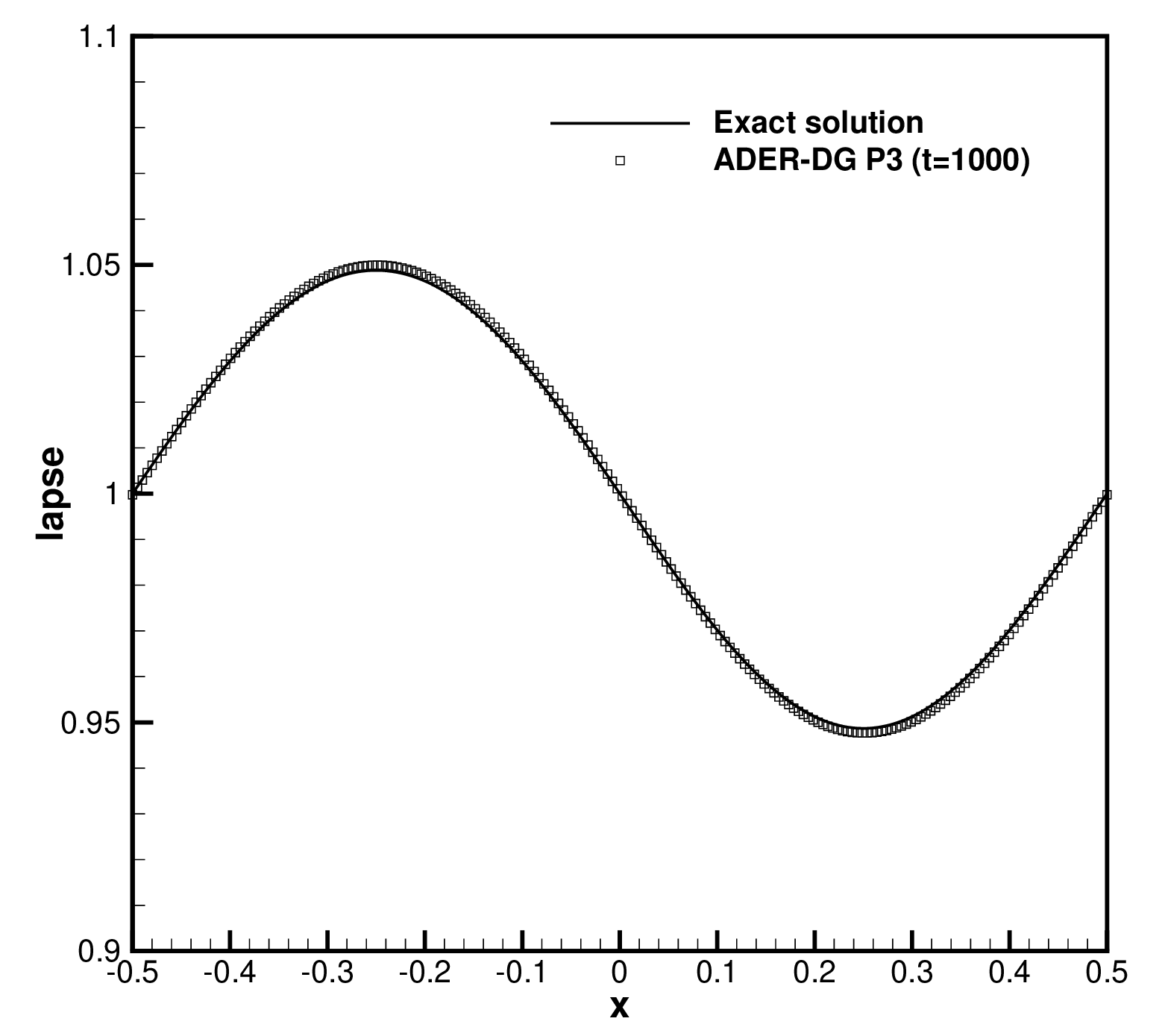}
\caption{\label{img.bench.gw}Large amplitude gauge wave}
\end{minipage} 
\end{figure}

Figure \ref{img.bench.bh} shows the Hamiltonian and momentum constraint violations during the evolution of a static Schwarzschild black hole spacetime with the FO-CCZ4 system. Limiting is applied only in the vicinity of the puncture. Figure \ref{img.bench.gw} shows a smooth gauge wave after 160 crossing times which is recovered with 100 cells at third order ($p=3$) already very good.

\section{Summary}\label{sec.summary}

We have shown first results on a new code which is the
unification of three different scientific branches: The
development of a new formulation of Einstein Equations, the
application of sophisticated high-order schemes and the
implementation in a pioneering high performance grid framework.

The proposed novel first order CCZ4 formulation was proven to be
strongly hyperbolic. Due to the special treatment of the
nonconservative terms as Borel measures, the system can be evolved very
accurately. The dynamical coupling of CCZ4 and GRMHD were sketched
and are subject to ongoing investigations.

The ADER-DG scheme provides high order time integration and
high order space discretization. The local space-time predictor
is communication avoiding and allows to integrate with only one
data transfer/grid traversal per time step \cite{exahype-review}.
The goal is to
maximize the arithmetic intensity (``science per watt''). The
subcell limiter preserves the degrees of freedom by resolving
physical oscillations with adaptive mesh refinement. It is shock
capturing as it allows to evolve shock fronts stably with any finite
volume scheme.

The Peano framework provides a modern AMR framework where users give up control in order to obtain scalability and maintainablity. Peano promises scaling to millions of cores. All together, the three ingredients (i) CCZ4+GRMHD (ii) the communication avoiding ADER-DG and (iii) Peano, were merged to a single
code called ExaHyPE from which first results have been shown in this work and more are expected soon.

\ack{S.K. thanks the members of the ExaHyPE consortium, especially Dominic Charrier, Francesco Fambri and Michael Dumbser, for contributing plots and results to this proceeding. The author also thanks Tobias Weinzierl, Michael Bader and especially Luciano Rezzolla for discussion and support. The author also
thanks the Helmholtz Graduate School HGS-HiRE for FAIR and the
Frankfurt International Graduate School for Science for support.
This project has received funding from the European Unions Horizon
2020 research and innovation programme under grant agreement  No 671698 (ExaHyPE).}

\section*{References}
\bibliography{references8}

\providecommand{\newblock}{}
\begin{thebibliography}{10}
\expandafter\ifx\csname url\endcsname\relax
  \def\url#1{{\tt #1}}\fi
\expandafter\ifx\csname urlprefix\endcsname\relax\def\urlprefix{URL }\fi
\providecommand{\eprint}[2][]{\url{#2}}

\bibitem{GW150914}
Abbott B~P {\em et~al.\/} (LIGO Scientific Collaboration and Virgo
  Collaboration) 2016 {\em Phys. Rev. Lett.\/} {\bf 116}(6) 061102
  \urlprefix\url{https://link.aps.org/doi/10.1103/PhysRevLett.116.061102}

\bibitem{GW170817}
Abbott B~P {\em et~al.\/} (LIGO Scientific Collaboration and Virgo
  Collaboration) 2017 {\em Phys. Rev. Lett.\/} {\bf 119}(16) 161101
  \urlprefix\url{https://link.aps.org/doi/10.1103/PhysRevLett.119.161101}

\bibitem{Pretorius:2005gq}
Pretorius F 2005 {\em Phys. Rev. Lett.\/} {\bf 95} 121101 (\textit{Preprint}
  \eprint{gr-qc/0507014})

\bibitem{Baiotti04}
Baiotti L, Hawke I, Montero P~J, L{\"o}ffler F, Rezzolla L, Stergioulas N, Font
  J~A and Seidel E 2005 {\em Phys. Rev. D\/} {\bf 71} 024035 (\textit{Preprint}
  \eprint{gr-qc/0403029})

\bibitem{Duez05MHD0}
Duez M~D, Liu Y~T, Shapiro S~L and Stephens B~C 2005 {\em Phys. Rev. D\/} {\bf
  72} 024028 astro-ph/0503420

\bibitem{Anninos05c}
Anninos P, Fragile P~C and Salmonson J~D 2005 {\em Astrophys. J.\/} {\bf 635}
  723

\bibitem{Anton06}
Ant{\'o}n L, Zanotti O, Miralles J~A, Mart{\'\i} J~M, Ib{\'a}{\~n}ez J~M, Font
  J~A and Pons J~A 2006 {\em Astrophys. J.\/} {\bf 637} 296 (\textit{Preprint}
  \eprint{astro-ph/0506063})

\bibitem{Giacomazzo:2007ti}
Giacomazzo B and Rezzolla L 2007 {\em Classical Quantum Gravity\/} {\bf 24}
  S235

\bibitem{Radice2012a}
{Radice} D and {Rezzolla} L 2012 {\em Astron. Astrophys.\/} {\bf 547} A26
  (\textit{Preprint} \eprint{1206.6502})

\bibitem{Radice2013b}
{Radice} D, {Rezzolla} L and {Galeazzi} F 2013  (\textit{Preprint}
  \eprint{arXiv:1306.6052})

\bibitem{Zanotti2015}
Zanotti O and Dumbser M 2015 {\em Computer Physics Communications\/} {\bf 188}
  110--127

\bibitem{Font08}
Font J~A 2008 {\em Living Rev. Relativ.\/} {\bf 6} 4
  \urlprefix\url{http://www.livingreviews.org/lrr-2008-7}

\bibitem{Marti2015}
{Mart{\'{\i}}} J~M and {M{\"u}ller} E 2015 {\em Living Reviews in Computational
  Astrophysics\/} {\bf 1}

\bibitem{exa1}
Kogge P, Bergman K, Borkar S {\em et~al.\/} 2008 Exascale computing study:
  Technology challenges in achieving exascale systems dARPA report
  \urlprefix\url{http://www.cse.nd.edu/Reports/2008/TR-2008-13.pdf}

\bibitem{Arnowitt62unfindable}
Arnowitt R, Deser S and Misner C~W 1962 {\em Gravitation: An introduction to
  current research\/} ed Witten L (New York: John Wiley) pp 227--265
  (\textit{Preprint} \eprint{gr-qc/0405109})

\bibitem{Baumgarte2010}
{Baumgarte} T~W and {Shapiro} S~L 2010 {\em {Numerical Relativity: Solving
  Einstein's Equations on the Computer}\/} (Cambridge University Press,
  Cambridge UK)

\bibitem{Gourgoulhon2012}
{Gourgoulhon} E 2012 {\em {3+1 Formalism in General Relativity}\/} ({\em
  Lecture Notes in Physics, Berlin Springer Verlag\/} vol 846)

\bibitem{Rezzolla_book:2013}
{Rezzolla} L and {Zanotti} O 2013 {\em {Relativistic Hydrodynamics}\/} (Oxford
  University Press, Oxford UK)

\bibitem{Shibata95}
Shibata M and Nakamura T 1995 {\em Phys. Rev. D\/} {\bf 52} 5428

\bibitem{Baumgarte99}
Baumgarte T~W and Shapiro S~L 1998 {\em Phys. Rev. D\/} {\bf 59} 024007

\bibitem{Bona:2003fj}
Bona C, Ledvinka T, Palenzuela C and Zacek M 2003 {\em Phys. Rev. D\/} {\bf 67}
  104005

\bibitem{Alic:2009}
{Alic} D, {Bona} C and {Bona-Casas} C 2009 {\em Phys. Rev. D\/} {\bf 79} 044026

\bibitem{Alic:2011a}
{Alic} D, {Bona-Casas} C, {Bona} C, {Rezzolla} L and {Palenzuela} C 2012 {\em
  Phys. Rev. D\/} {\bf 85}(6) 064040

\bibitem{Alic2013}
{Alic} D, {Kastaun} W and {Rezzolla} L 2013  (\textit{Preprint}
  \eprint{1307.7391})

\bibitem{ADERCCZ4}
Dumbser M, Guercilena F, K\"oppel S, Rezzolla L and Zanotti O 2017 {\em
  submitted to Physical Review D\/} (\textit{Preprint} \eprint{1707.09910})

\bibitem{NGMD2006}
{Noble} S~C, {Gammie} C~F, {McKinney} J~C and {Del Zanna} L 2006 {\em Astrop.
  J.\/} {\bf 641} 626--637

\bibitem{ZannaZanotti}
Del~Zanna L, Zanotti O, Bucciantini N and Londrillo P 2007 {\em Astronomy and
  Astrophysics\/} {\bf 473} 11--30

\bibitem{BHAC2017}
Porth O, Olivares H, Mizuno Y, Younsi Z, Rezzolla L, Moscibrodzka M, Falcke H
  and Kramer M 2017 {\em Computational Astrophysics and Cosmology\/} {\bf 4} 1

\bibitem{fambri17}
Fambri F, Dumbser M, K\"oppel S, Rezzolla L and Zanotti O 2018 {\em Monthly
  Notices of the Royal Astronomical Society\/} {\bf 477} 4543--4564
  \urlprefix\url{http://dx.doi.org/10.1093/mnras/sty734}

\bibitem{Zanotti2015c}
Zanotti O, Fambri F, Dumbser M and Hidalgo A 2015 {\em Computers and Fluids\/}
  {\bf 118} 204 -- 224 ISSN 0045-7930

\bibitem{Zanotti2015d}
Zanotti O, Fambri F and Dumbser M {2015} {\em Mon. Not. R. Astron. Soc.\/} {\bf
  452} 3010--3029

\bibitem{ADERDGVisc}
Fambri F, Dumbser M and Zanotti O 2017 {\em Computer Physics Communications\/}
  ISSN 0010-4655
  \urlprefix\url{http://www.sciencedirect.com/science/article/pii/S0010465517302448}

\bibitem{NCP_HLLEM}
Dumbser M and Balsara D~S 2016 {\em Journal of Computational Physics\/} {\bf
  304} 275--319 ISSN 0021-9991

\bibitem{Dumbser2008}
{Dumbser} M, {Balsara} D~S, {Toro} E~F and {Munz} C~D 2008 {\em Journal of
  Computational Physics\/} {\bf 227} 8209--8253

\bibitem{ADERNC}
Dumbser M, Castro M, {Par\'es} C and Toro E 2009 {\em Computers and Fluids\/}
  {\bf 38} 1731--1748

\bibitem{exahype-review}
{Charrier} D~E and {Weinzierl} T 2018 Stop talking to me -- a
  communication-avoiding ader-dg realisation {\emph{in preparation}}

\bibitem{loeffler_2011_et}
L{\"{o}}ffler F, Faber J, Bentivegna E, Bode T, Diener P, Haas R, Hinder I,
  Mundim B~C, Ott C~D, Schnetter E, Allen G, Campanelli M and Laguna P 2011
  {\em ArXiv e-prints\/} (\textit{Preprint} \eprint{1111.3344})

\bibitem{Goodale02a}
Goodale T, Allen G, Lanfermann G, Mass{\'o} J, Radke T, Seidel E and Shalf J
  2003 {\em Vector and Parallel Processing -- VECPAR'2002, 5th International
  Conference, Lecture Notes in Computer Science\/} (Berlin: Springer)

\bibitem{loeffler2013}
L{\"{o}}ffler F, Brandt S~R, Allen G and Schnetter E 2013 {\em CoRR\/} {\bf
  abs/1309.1812} \urlprefix\url{http://arxiv.org/abs/1309.1812}

\bibitem{schnetter14}
Schnetter E, Blazewicz M, Brandt S~R, Koppelman D~M and L{\"{o}}ffler F 2014
  {\em CoRR\/} {\bf abs/1410.1764}
  \urlprefix\url{http://arxiv.org/abs/1410.1764}

\bibitem{exahype-web}
{Charrier} D~E, {Dumbser} M, {Duru} K, {Fambri} F, {Gabriel} A, {Gallard} J~M,
  {K\"oppel} S, {Osorio} A, {Rannabauer} L, {Rezzolla} L, {Tavelli} M and
  {Weinzierl} T 2017 Exahype, an exascale hyperbolic pde engine
  \url{http://www.exahype.eu}

\bibitem{exahype-guidebook}
{Schwarz} A, {Charrier} D~E, {Guera} F, {Gallard} J~M, {Samfass} P, {K\"oppel}
  S, {Weinzierl} T and {Varduhn} V 2017 {The ExaHyPE guidebook}
  \url{http://dev.exahype.eu/guidebook.pdf}

\bibitem{Weinzierl15}
Weinzierl T 2015 {\em CoRR\/} {\bf abs/1506.04496}
  \urlprefix\url{http://arxiv.org/abs/1506.04496}

\bibitem{bader-sfc}
Bader M 2013 {\em {Space-Filling Curves: An Introduction with Applications in
  Scientific Computing}\/} (Springer)

\bibitem{sod}
Sod G~A 1978 {\em Journal of Computational Physics\/} {\bf 27} 1--31

\bibitem{toro-book}
Toro E 1999 {\em {{R}iemann} Solvers and Numerical Methods for Fluid
  Dynamics\/} 2nd ed (Springer)

\bibitem{Balsara1999b}
{Balsara} D~S and {Spicer} D 1999 {\em Journal of Computational Physics\/} {\bf
  148} 133--148

\bibitem{Font02a}
Font J~A and Daigne F 2002 {\em Mon. Not. R. Astron. Soc.\/} {\bf 334} 383--400

\bibitem{Tolman39}
Tolman R~C 1939 {\em Phys. Rev.\/} {\bf 55} 364

\end{thebibliography}

\end{document}